\documentstyle[epsf,prl,preprint,aps]{revtex}
\tightenlines
\begin{document}
\draft
\title{Ultraslow dynamics and stress relaxation in the aging of a soft glassy system}
\author{Laurence Ramos and  Luca Cipelletti}
\address{Groupe de Dynamique des Phases Condens\'{e}es (UMR 5581),
Universit\'{e} Montpellier 2, 34095 Montpellier Cedex 5, France}
\date{\today}
\maketitle
\begin{abstract}

We use linear rheology and multispeckle dynamic light scattering
(MDLS) to investigate the aging of a gel composed of multilamellar
vesicles. Light scattering data indicate rearrangement of the gel
through an unusual ultraslow ballistic motion. A dramatic slowdown
of the dynamics with sample age $t_{w}$ is observed for both
rheology and MDLS, the characteristic relaxation time scaling as
$t_{w}^{\mu}$. We find the same aging exponent $\mu =0.78$ for
both techniques, suggesting that they probe similar physical
processes, that is the relaxation of applied or internal stresses
for rheology or MDLS, respectively. A simple phenomenological
model is developed to account for the observed dynamics.

\end{abstract}

The fascinating structural properties of colloidal systems are now
richly documented experimentally and, for most, theoretically well
understood. Modeling their dynamics remains however more
challenging. In particular, the extremely slow dynamics and aging
properties of disordered, out-of-equilibrium systems raise
puzzling questions. A theoretical approach has been recently
proposed by Sollich \textit{et al.} \cite{Sollich1,Sollich2}, who
introduced the notion of soft glassy materials and modeled the
glass-like transition and the linear and nonlinear rheology
properties of out-of-equilibrium soft systems, with a theory
adapted from glasses. Despite the large variety of materials which
belong to the soft glassy materials classification (dense
emulsions, pastes, foams, gels, ...), the experimental
investigations of their out-of-equilibrium properties still remain
scarce, mostly because of experimental difficulties.

In this Letter, we study close-packed multilamellar vesicles
(MLVs). These systems are typically formed by shearing a
surfactant lamellar phase, otherwise relatively flat, above a
critical shear rate \cite{onions}. Here, a compact arrangement of
MLVs is obtained by incorporating an amphiphilic block-copolymer
into the surfactant lamellar phase \cite{onions2}. We show that
this system is metastable, as evidenced by the evolution of its
dynamics and mechanical properties and we present an experimental
investigation of the aging process of this material by means of
two independent techniques: dynamic light scattering (DLS) and
linear rheology. Both the rheological response function (stress
relaxation following a step-strain) and the intensity
autocorrelation function, measured by DLS, exhibit a dramatic
slowdown of their characteristic relaxation time $\tau$ with
sample age $t_{w}$. In both cases, $\tau$ increases as a powerlaw
with age: $\tau \sim t_{w}^{\mu}$. Remarkably, we find the same
aging exponent $\mu = 0.78$ for both rheology and DLS, thus
strongly suggesting that the two techniques probe the same
physical process. However, an intriguing difference exists in the
shape of the relaxation functions. The rheology relaxation
function is a stretched exponential with a small stretching
exponent $m\approx 0.2$, while the correlation function measured
by DLS displays an unusually large stretching exponent $p \approx
1.5$, associated with a $1/q$ dependence of the relaxation time,
where $q$ is the scattering vector. We discuss these results in
terms of rearrangement of the texture of the sample driven by the
relaxation of internal and applied stresses through ballistic
motion of the MLVs.

\begin{figure}
\centerline{\epsffile{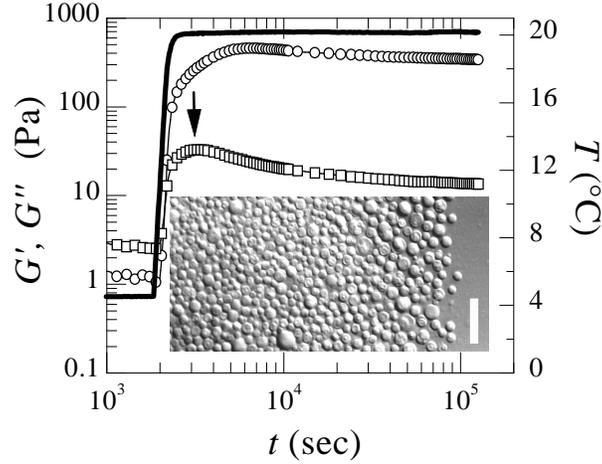}}
 \caption{Time and temperature
dependency of the storage $G'$ (circles) and loss $G"$ (squares)
moduli of a lamellar gel ($\Omega = 0.8$) (frequency $1 \,
\rm{Hz}$, strain amplitude $1 \%$); at $t=1800 \, \rm{sec}$ the
temperature is increased from $4.6$ to $20 \, ^{\circ}\rm{C}$
(solid line). Age $t_{w}=0$ of the gel is taken at the maximum of
$G"$ (arrow). Inset : Optical microscopy picture showing the
interface between the lamellar gel (left) and the solvent (right).
Temperature is $20 \, ^{\circ}\rm{C}$ and scale bar is $20 \,
\rm{\mu m}$.} \label{Fig1:1}
\end{figure}

The lamellar phase is constituted by bilayers composed of a
mixture of cetylpyridinium chloride (CpCl) and octanol (Oct)
(weight ratio $\rm{CpCl/Oct}=0.95$), diluted in brine
($\rm{[NaCl]}= 0.2 \rm{M}$) at a weight fraction of $16 \%$. The
smectic periodicity of the lamellae is $16 \,\rm{nm}$, as measured
by neutron scattering \cite{Ligoure}. The bilayers are decorated
by an amphiphilic copolymer, Symperonics F68 by Serva
($\rm{(EO)_{76}}-\rm{(PO)_{29}}-\rm{(EO)_{76}}$, where EO is
ethylene oxide and PO is propylene oxide). The
copolymer-to-bilayer weight ratio $\Omega$ ranges from $0$ to
$1.6$. Upon copolymer addition, the lamellar structure is
preserved but a marked and continuous hardening of the system is
observed, resulting in a lamellar gel \cite{Ligoure,Safinya}.
Polarized light microscopy observation of the gel reveals a
texture characteristics of a dense arrangement of MLVs
\cite{onions}. Additional evidence for the large-scale structure
of the bilayers is provided by microscope images of a gel brought
in contact with the solvent: as shown in the inset of
Fig.~\ref{Fig1:1}, detachment of individual MLVs close to the
gel/solvent boundary is observed, whose size distribution is
relatively narrow (diameter $\approx 5.0 \pm 0.8 \, \mu \rm{m}$).
In the gel, on the contrary, MLVs have polyhedral shape; a
three-dimensional network of flat interfaces, similar to that of a
dried foam, separates domains where the lamellae have the same
orientation. Refractive index fluctuations at these flat
interfaces scatter light; indeed, the light scattered intensity
decays as $q^{-3.76 \pm 0.03}$ for $3 < q < 31
 \, \mu\rm{m}^{-1}$, a powerlaw behavior characteristic of the
scattering from a disordered network of planes \cite{Leng}.

We perform both oscillatory and stationary rheological experiments
in a Couette geometry with a Paar Physica UDS 200
stress-controlled rheometer, operated in the strain-controlled
mode through a feedback loop. All measurements are done in the
linear regime. The temperature and time dependency of the elastic
modulus, $G'$, and loss modulus, $G"$, of a gel are reported in
Fig.~\ref{Fig1:1}. At a temperature $T=4 \, \rm{^{o}C}$ the
material is a viscoelastic fluid, characterized by a weak complex
modulus of the order of $1 \, \rm{Pa}$ whose imaginary part is
dominant. A temperature ramp to $T=20 \, \rm{^{o}C}$ increases by
several orders of magnitude both moduli. The system presents
thereafter the characteristic of a gel : the elastic modulus is
about one order of magnitude larger than the loss modulus and is
nearly frequency-independent in the range $10^{-2}-10 \,\rm{Hz}$.
The gel thus formed is an out-of-equilibrium system, as shown by
the time evolution of $G'$ and $G"$. Indeed, after the formation
of the gel both moduli evolve continuously, never reaching a
steady value. Figure~\ref{Fig1:1} shows the rapid increase of $G'$
and $G"$ during the temperature ramp, followed by their slow but
continuous decrease with time, over a period of $2$ days.
Interestingly, the time dependence of the moduli is dominated by
the variation of the loss modulus, whose relative decrease is
always larger than that of $G'$, consistently with theoretical
predictions for the aging of soft glassy materials
\cite{Sollich2}. The temperature dependence of the complex modulus
(Fig.~\ref{Fig1:1}) is fully reversible. Thus, the thermal
treatment allows us to erase any memory of the loading procedure
and to perform measurements from a well controlled initial state.
We define age $t_w = 0$ at the maximum of $G"$ (arrow in
Fig.~\ref{Fig1:1}).

\begin{figure}
\centerline{\epsffile{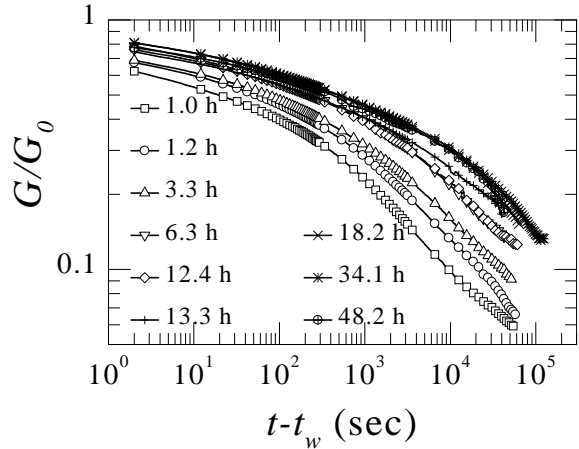}}
 \caption{Relaxation modulus
$G(t,t_{w})$ following a step strain of amplitude $1\%$ of a
lamellar gel ($\Omega = 0.8$). For the sake of clarity,
$G(t,t_{w})$ has been normalized by its value $G_{0}$ at time
$t-t_{w}=0$. Curves are labeled by sample age $t_{w}$.}
 \label{Fig2:2}
\end{figure}

\begin{figure}
\centerline{\epsffile{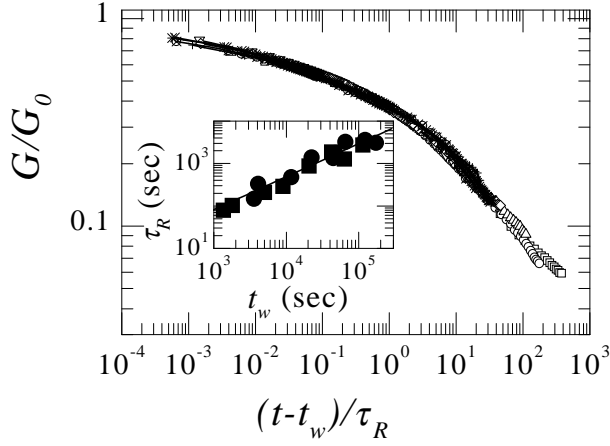}} \caption{Scaling of $G/G_{0}$
versus $(t-t_{w})/\tau_{R}$ for the data of Fig.~\ref{Fig2:2}.
Inset: Age dependence of the relaxation time $\tau_{R}$ for a gel
with $\Omega = 0.8$ (circles) and $\Omega = 1.2$ (squares). The
straight line is a power law fit yielding an exponent of $0.78$.}
 \label{Fig3:3}
\end{figure}

The typical time evolution of the linear step-strain response
function, $G(t,t_{w})$, is shown in Fig.~\ref{Fig2:2}, for a
strain amplitude $\gamma  = 1 \, \%$ \cite{Linear}. The different
curves in Fig.~\ref{Fig2:2} correspond to different ages $t_{w}$
of the sample: after forming the gel, the sample ages unperturbed
during a waiting time $t_{w}$ before a step strain is applied. In
all cases, on time scales of several hours, a full decay of the
stress is obtained. The whole relaxation curve is well described
by a stretched exponential function,
$G(t,t_{w})=G_{0}\exp[-((t-t_{w})/\tau_{R})^{m}]$, where
$\tau_{R}$ and $m$ are the relaxation time and the stretching
exponent for rheology, respectively. The data exhibit a dramatic
slowdown of the relaxation with sample age, the characteristic
time $\tau_{R}$ increasing as a powerlaw of $t_{w}$: $\tau_{R}
\sim t_{w}^{\mu}$, with $\mu=0.78 \pm 0.09$. However, despite the
marked increase of the characteristic time, the relaxation
functions measured at different ages have all the same shape, and
are thus characterized by the same age-independent stretching
exponent $m$, as demonstrated by the collapse of data taken at
different ages onto a master curve when $G/G_{0}$ is plotted
against $(t-t_{w})/\tau_{R}$ (Fig.~\ref{Fig3:3}). For the sample
$\Omega=0.8$ shown in Figs.~\ref{Fig2:2} and~\ref{Fig3:3},
$m=0.19\pm 0.01$. Qualitatively similar master curves have been
obtained with different soft disordered systems \cite{Cloitre}.

\begin{figure}
\centerline{\epsffile{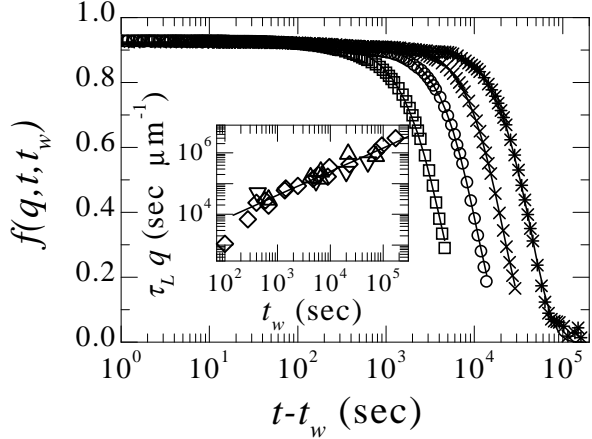}} \caption{Dynamic structure
factors at $q=11.2 \, \mu\rm{m}^{-1}$ for a sample with $\Omega =
0.8$. Sample age is $7 \, \rm{min}$ (squares), $1.5 \, \rm{h}$
(circles), $5.6 \, \rm{h}$ (crosses) and $14.3 \, \rm{h}$ (stars).
Solids lines are best fits using a stretched exponential function.
Inset: Age dependence of $q\tau_{L}$, where $\tau_{L}$ is the
characteristic time derived from MDLS and $q=6.0$ (up triangles),
$11.2$ (down triangles) and $24.2 \, \mu\rm{m}^{-1}$ (diamonds).
The straight line is a power law fit yielding an exponent of
$0.77$.}
 \label{Fig4:4}
\end{figure}

To independently probe the aging of the MLVs gels, we use a
completely different technique, multispeckle dynamic light
scattering (MDLS) \cite{Luca2D}. A charge-coupled device camera is
used as a multi-element detector in place of a phototube on a
standard light scattering apparatus. Intensity autocorrelation
functions are averaged over several thousands pixels,
corresponding to about 500 independent speckles or areas of
coherence associated to a small solid angle centered around the
direction of a given scattering vector $\mathbf{q}$ (the spread in
scattering vector is $\Delta q/q \simeq 10^{-2}$). Thanks to the
multispeckle technique, the typical measurement duration is
reduced to the longest delay in the correlation function, thus
allowing snapshots of the dynamics to be taken, thereby directly
probing the aging. Samples are initialized by a thermal treatment
equivalent to that performed in the rheology experiments. Figure
\ref{Fig4:4} shows the time evolution of the dynamic structure
factor $f(q,t,t_w)$, measured at $q=11.2 \, \mu\rm{m}^{-1}$.
Except for the initial decay of $f(q,t,t_w)$, whose characteristic
time (of the order of $0.5 \, \rm{msec}$) is too fast to be
measured by MDLS and thus is not visible in Fig. \ref{Fig4:4}, the
dynamics is essentially frozen on time scales up to several
thousands seconds. Surprisingly, although the system is solid, a
full relaxation of $f(q,t,t_w)$ is eventually observed, indicating
rearrangements on length scales up to $1 \, \mu \rm{m}$. The
relaxation characteristic time increases with age, reaching values
as large as $10^{5} \, \rm{sec}$. Similarly to the rheology
response functions, the final relaxation of MDLS correlation
functions is well fitted by a stretched exponential:
$f(q,t,t_w)=A\exp[-((t-t_{w})/\tau_{L})^{p}]$, where $\tau_{L}$
depends on both $q$ and $t_{w}$. However, the shape of the
correlation functions is very different from that of $G(t,t_{w})$,
since for MDLS the stretching exponent $p$ is always larger than
$1$: $p=1.46\pm 0.18$. Such an unusual decay has recently been
observed for colloidal gels \cite{Lucagel} and was associated with
a very peculiar $1/q$ dependence of the relaxation time.
Remarkably, the same dependence is found here, as demonstrated by
a plot of $\tau_{L}q$ versus age, where data taken at different
scattering vectors collapse onto the same master curve (inset of
Fig.~\ref{Fig4:4}). Aging leads to an increase by more than three
orders of magnitude of the relaxation time $\tau_{L}$, which
scales as $\tau_{L} \sim t_{w}^{\mu}$, with $\mu=0.77 \pm 0.04$.
Strikingly, we find the same age dependency for the relaxation
time probed by MDLS and rheology, since both $\tau_{L}$ and
$\tau_{R}$ increase as a powerlaw with $t_w$, with, within
experimental errors, the same aging exponent $\mu$.

Aging indicates that a significant part of the relaxation takes
place on time scales close to the sample age, leading to a nearly
linear increase of the characteristic relaxation time with age, a
feature common to several glassy materials
\cite{Cloitre,Lucagel,Struik,Laponite}. We note that, on the time
scale of the experiments presented here, no modification of the
structure of the MLVs is observed. Therefore we exclude structural
evolution as a driving force for aging. Instead, we propose that
the dynamics be due to rearrangements of the texture, driven by
the minimization of internal stresses. These rearrangements lead
to a hardening of the system, as measured by rheology. Moreover,
we point out that internal stresses were found to be responsible
for the same unusual dynamics in colloidal gels \cite{Lucagel};
they certainly exist also in the glassy state of a compact
arrangement of soft objects and correspond to excess of stored
elastic energy quenched in the disordered structure
\cite{Internstress}. Further support for our interpretation comes
from the fact that identical aging processes are found by MDLS and
rheology, where the relaxation of an applied stress is directly
measured.

A distinctive feature of the slow dynamics reported here is that
the final decay of the dynamic structure factor $f(q,t,t_{w})$
depends uniquely on the product $x=q(t-t_{w})$. This is the
hallmark of ballistic motion \cite{ballistic}, with a
characteristic velocity $\bar{V} = (q\tau_{L})^{-1}$.  The
velocity distribution function of the scatterers can be obtained
from the shape of $f(x)$. In the case of isotropic motion, the
probability $W(V)$ for the velocity modulus to lie between $V$ and
$V+dV$ reads \cite{ballistic} :

\begin{equation}
W(V)=\frac{2V}{\pi} \int_{0}^{+ \infty} dx \, x \, f(x) \, sin(xV)
\label{Eq1}
\end{equation}
In our case, $f(q,t,t_{w})=f(x) \sim \exp[-(x\bar{V})^{p}]$ is
proportional to the Fourier transform of the Levy stable law
\cite{Levy}, $L_{p,0}(V/\bar{V})=\frac{\bar{V}}{\pi}\int_{0}^{+
\infty} dx \, \cos(xV) \, \exp[-(x\bar{V})^{p}]$. Thus, one
obtains for the velocity distribution function :

\begin{equation}
W(V) = -2V/\bar{V} \frac{d}{dV}[L_{p,0}(V/\bar{V})] \label{Eq2}
\end{equation}
The Levy law is characterized by a broad tail :
$L_{p,0}(V/\bar{V}) \sim V^{-(p+1)}$ for $V \gg \bar{V}$; Eq.
~\ref{Eq2} therefore predicts asymptotically $W(V) \sim V^{-(p+1)}
= V^{-(2.46\pm 0.18)}$.

Simple arguments may account for this powerlaw dependence. The
deformation field $\Delta R (r,t)$ at a distance $r$ from a source
of stress is that of an elastic solid. In the absence of external
stress sources,  the leading (dipolar) term of a multipolar
expansion of $\Delta R$ is  $\alpha(t)r^{-2}$. Because of the
relatively high elastic modulus of the material, the propagation
of elastic stresses is almost instantaneous when compared to
$\tau_{L}$. Thus, the slow dynamics cannot be due to the
propagation of the elastic deformation but rather to the time
evolution of the strength of the internal stress sources.
Therefore, the ballistic motion found by MDLS indicates $\alpha(t)
\sim t$ and $V(r) = dR/dt \sim r^{-2}$. Note that a linear
increase with time of the  strain has been previously assumed in a
phenomenological model for fractal gels \cite{Lucagel} and is also
obtained by a more sophisticated model \cite{Pitard}. For randomly
distributed stress sources, the number of such sources at a
distance between $r$ and $r+dr$ from any given point of the gel is
$dN \sim r^{2}dr$. Therefore
 $W(V) \sim dN/dV \sim r^{5} \sim V^{-\rm{2.5}}$, in very good
agreement with the experimental results. Note that aging indicates
that the internal stresses decrease with time, because either
their strength $\alpha$ or the number of sources $N$ decreases.

The ballistic-motion rearrangement of the MLVs is also presumably
responsible for the linear stress relaxation measured by rheology.
The time scales measured by rheology and MDLS are however quite
different. In fact, one might expect that a mechanical stress is
relaxed when MLVs have moved over a distance $l$ comparable to
their size ($l \approx 5 \, \mu \rm{m}$). The rheology relaxation
time is then estimated to be of order of $l/\bar{V}$. However,
such value is more than three orders of magnitude larger than
$\tau_{R}$. This discrepancy underlines the fundamental difference
between a DLS experiment and a rheology one. Although both probe
the rearrangement of the MLVs, the two techniques are
intrinsically measuring different quantities. In fact, $\tau_{L}$
corresponds to the characteristic time for all scatterers to move
over a distance $1/q$. By contrast, in rheology, rearrangements of
a few MLVs may be sufficient to relax the applied stress;
$\tau_{R}$ is thus expected smaller than $\tau_{L}$, as observed.
In concentrated emulsions, it has indeed been measured that the
motion of only a few percent of the particles is sufficient for
the material to yield \cite{Echo}. Moreover, the fact that the
ratio $\tau_{R}/\tau_{L}$ is constant suggests that the fraction
of rearrangements required for yielding does not vary with sample
age. Finally, we observe that although the stretching exponents
$m$ and $p$ for rheology and MDLS are very different, they both
reflect very broad distributions of times and velocities,
respectively. To quantitatively relate the two distributions, more
information about the fraction of rearrangements required for the
material to yield would be needed.

The very peculiar time-dependent slow dynamics reported here
should be observable with other repulsive systems, such as dense
emulsions or foams, whose elasticity shares many features with
that of MLVs gels. Moreover, the intriguing analogies found with
colloidal gels \cite{Lucagel}, an attractive system, and with a
micellar polycrystal \cite{MicellarPolyXtal} hint at a possible
universality of such behavior in jammed soft materials. More
experiments are needed to test this hypothesis.

We thank R. C. Ball, E. Pitard, J.-P. Bouchaud, W. Kob and G.
Porte for very useful discussions and suggestions and F.
Castro-Roman for initiating the rheology experiments.

%


\begin{references}
\bibitem{Linear} No changes in $G(t,t_{w})$
are observed when reducing $\gamma$, thus confirming that the data
are taken in the linear regime and showing that the linear regime
is essentially frequency-independent, as predicted for soft glassy
materials \cite{Sollich1}.
\end{references}
\end{document}